\documentclass[a4paper,twocolumn,3p,preprint]{elsarticle}

\usepackage{lineno,hyperref}
\modulolinenumbers[5]

\journal{Carbon}

\begin{document}

\title{Thermal Conductivity and Mechanical Properties of Nitrogenated Holey Graphene}
\author[buw]{Bohayra Mortazavi\corref{cor1}}
\ead{bohayra.mortazavi@gmail.com}

\author[buw]{Obaidur Rahaman}

\author[buw]{Timon Rabczuk}

\author[ufrn]{Luiz Felipe C. Pereira\corref{cor1}}
\ead{pereira@dfte.ufrn.br}

\address[buw]{Institute of Structural Mechanics, Bauhaus-Universit\"at Weimar, Marienstr. 15, D-99423 Weimar, Germany}
\address[ufrn]{Departamento de F\'{\i}sica Te\'orica e Experimental, Universidade Federal do Rio Grande do Norte, Natal, 59078-970, Brazil}

\cortext[cor1]{Corresponding author}

\date{\today}

\begin{abstract}
Nitrogenated holey graphene (NHG), a two-dimensional graphene-derived material with a C$_2$N stoichiometry and evenly distributed holes and nitrogen atoms in its basal plane, has recently been synthesized. We performed first principles calculations and molecular dynamics simulations to investigate mechanical and heat transport properties of this novel two-dimensional material at various temperatures. First principles calculations based on density functional theory yield an elastic modulus of $400 \pm 5$ GPa at 0 K, $10$\% larger than predicted by molecular dynamics simulations at low temperatures.
We observed an overall decreasing trend in elastic modulus and tensile strength as temperature increases. 
At room temperature, we found that NHG can present a remarkable elastic modulus of $335 \pm 5$ GPa and tensile strength of $60$ GPa. 
We also investigated the thermal conductivity of NHG via non-equilibrium molecular dynamics simulations.
At $300$ K an intrinsic thermal conductivity of $64.8$ W/m-K was found, with an effective phonon mean free path of $34.0$ nm, both of which are smaller than  respective values for graphene, and decrease with temperature.
Our modeling-based predictions should serve as guide to experiments concerning physical properties of this novel material.

\end{abstract}


\maketitle

\section{Introduction}

Since graphene's first reported observation \cite{Novoselov2004, Geim2007}, the family of two-dimensional (2D), atomically thin materials has grown considerably \cite{Bhimanapati2015}.
Materials such as silicene\cite{Lalmi2010}, germanene \cite{Bianco2013}, phosphorene \cite{Li2014a}, stanene \cite{Zhu2015b} and borophene \cite{Mannix2015}, to name just a few, have either been produced or isolated in single-layer or few-layer form.
Together with graphene, these materials present a complete range of electronic properties, including metals, semiconductors and insulators.
Regarding mechanical properties, most of these materials are quite resistant to tensile strain while presenting almost no resistance to compressive strain, due to their sheet-like structure.
Meanwhile, concerning heat transport properties, these 2D wonder-materials, also cover a wide range of values from low to very high.

\begin{figure}[htbp]
\begin{center}
\includegraphics[width=\linewidth]{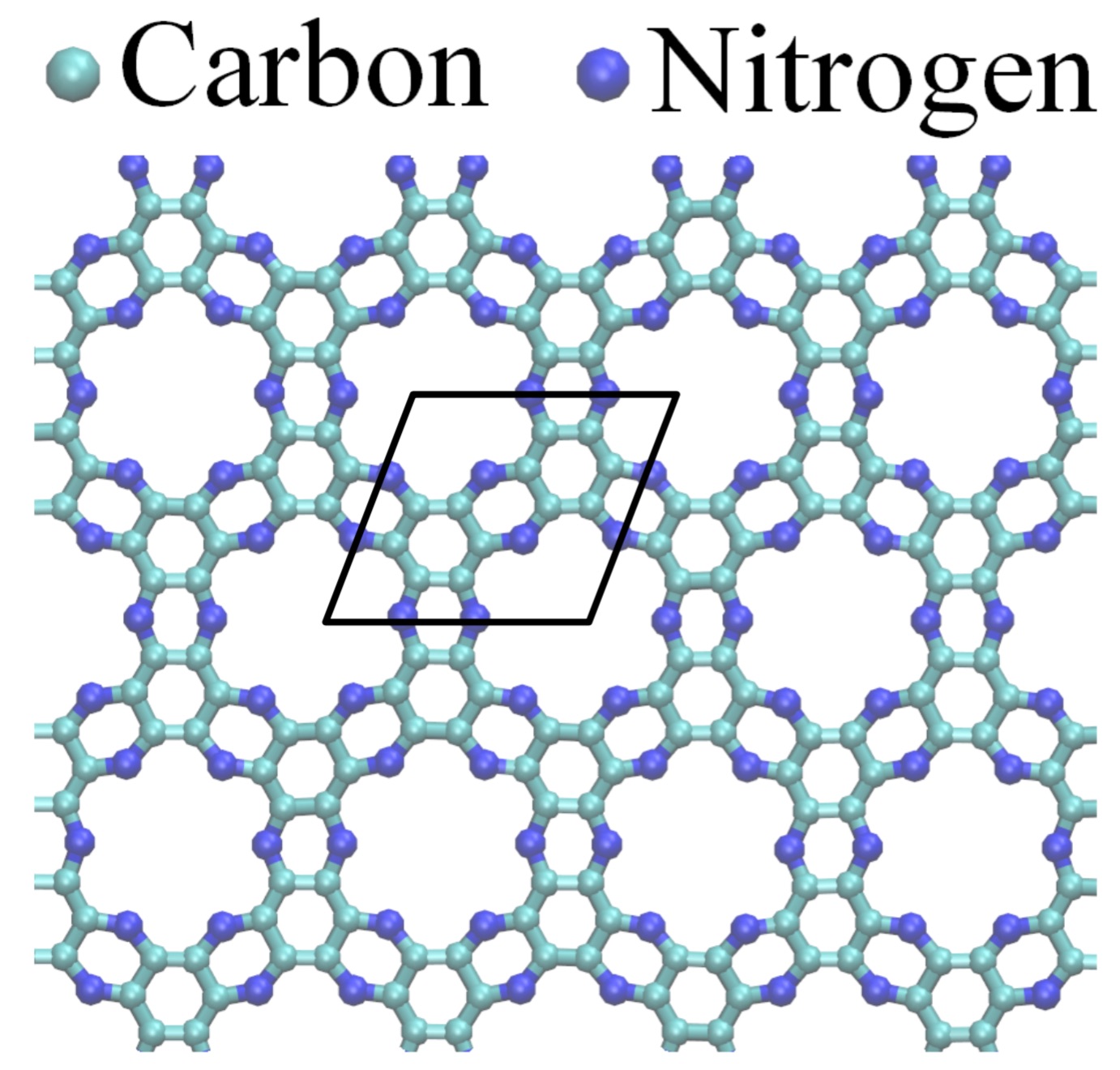}
\caption{Atomic structure of nitrogenated holey graphene, which consists of rings of carbon atoms terminated by nitrogen atoms. An 18-atoms unit cell is also shown.}
\label{fig1}
\end{center}
\end{figure}

Recently, yet another 2D material has been synthesized via direct functionalization of graphene sheets ~\cite{Mahmood2015}. This novel graphene-derived material presents a crystal structure with evenly distributed holes and nitrogen atoms with a C$_2$N stoichiometry in its basal plane. The atomic structure of nitrogenated holey graphene (NHG) consists of rings of carbon atoms terminated by nitrogen atoms, as show in Fig.~\ref{fig1}.
In this work we employed  extensive first principles calculations and classical molecular dynamics simulations in order to predict mechanical and heat transport properties of this novel material.
To the best of our knowledge there are no experimental nor theoretical investigations of such properties yet, therefore predictions based on our modeling results can serve as a guide to groups interested in the physical properties of this novel 2D material.

\section{Methods}

We have performed atomistic modeling of NHG via classical molecular dynamics simulations (MD) where the Tersoff bond-order potential was employed to describe interatomic forces~\cite{Tersoff1988, Tersoff1988a}.
Although the Tersoff potential has not been parameterized to model NHG specifically, it has already been parameterized for graphene~\cite{Lindsay2010}, hexagonal boron nitride (hBN)~\cite{Lindsay2011a}, and graphene--hBN structures~\cite{Kinaci2012}.
Therefore, in the present work we employed potential parameters from Lindsay and Broido~\cite{Lindsay2010,Lindsay2011a} for C-C interaction and obtained the parameters for C-N via the usual mixing rules for the Tersoff potential \cite{Tersoff1988a}.

\begin{figure}[htbp]
\begin{center}
\includegraphics[width=\linewidth]{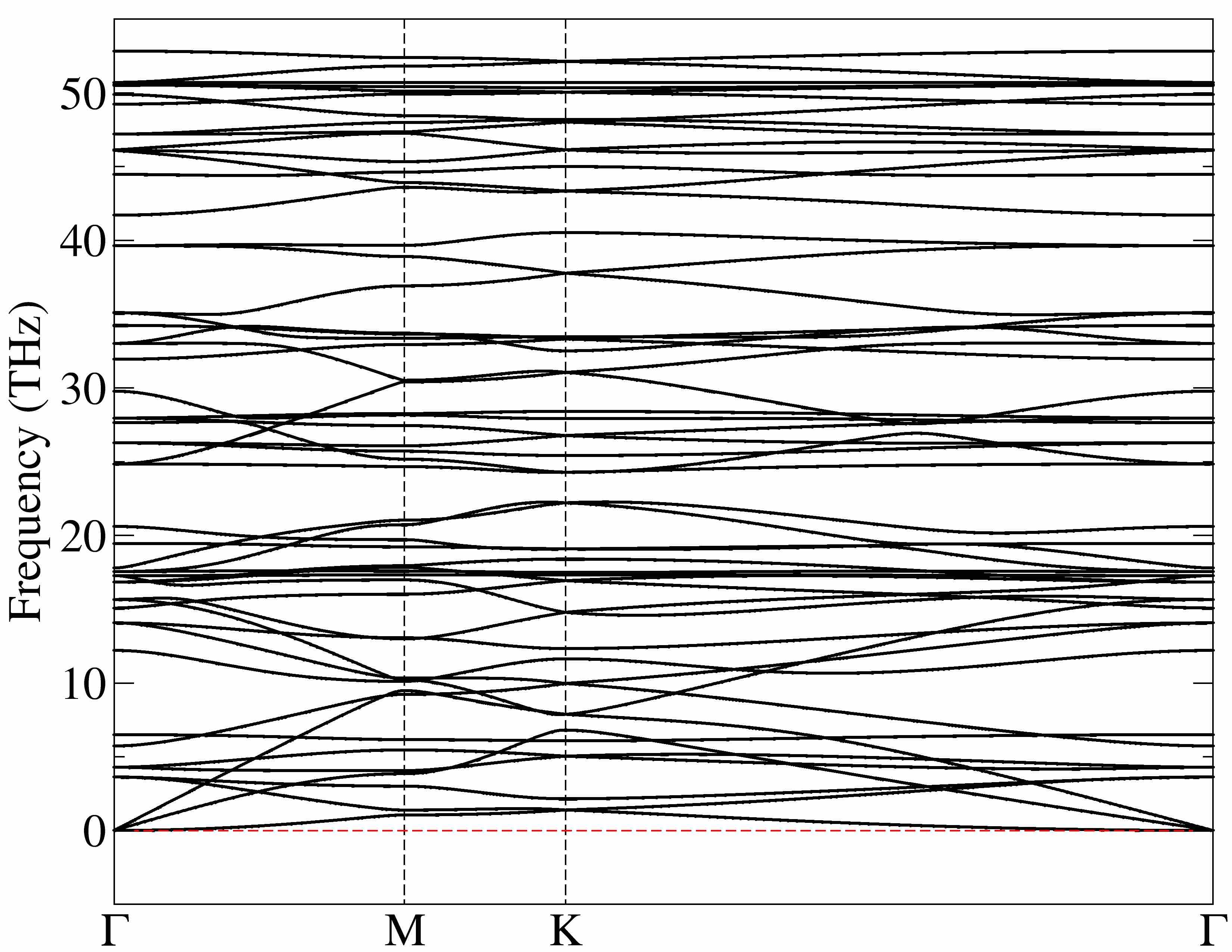}
\caption{Phonon dispersion of NHG. The absence of imaginary frequencies indicates the stability of this crystal structure with the chosen potential parameters.}
\label{fig:pdisp}
\end{center}
\end{figure}

In order to verify the accuracy of the chosen potential parameters, we first obtained phonon dispersion relations of NHG by diagonalizing the dynamical matrix via the General Utility Lattice Program (GULP)~\cite{Gale1997, Gale2003}.
The phonon dispersion presented in Fig.~\ref{fig:pdisp} was obtained for an 18--atom unit cell (shown in Fig.~\ref{fig1}). 
From the dispersion curves we see that NHG presents three acoustic modes: two of the modes present linear dispersion around the $\Gamma$-point, while the third mode has a quadratic dispersion, similar to graphene and hBN~\cite{Lindsay2010, Lindsay2011a}. 
The absence of imaginary frequencies in Fig. ~\ref{fig:pdisp} indicates that the crystal structure is stable when described by the chosen potential parameters, at least at $0$ K.
Furthermore, the features observed in Fig.~\ref{fig:pdisp} are, at least qualitatively, in agreement with recent ab-initio calculations based on density functional theory (DFT), which also confirmed the stability of NHG's predicted crystal structure~\cite{Sahin2015}.

We also investigated the mechanical response of NHG sheets by MD simulations and first principles DFT calculations. 
MD simulations were performed with the Large-scale Atomic/Molecular Massively Parallel Simulator (LAMMPS)\cite{Plimpton1995} package, with periodic boundary conditions along in-plane directions and free boundary conditions in the perpendicular direction. 
Mechanical properties were evaluated by performing uniaxial tensile strain along in-plane directions.
Each simulation cell was equilibrated at the desired temperature via a Nos\'e-Hoover thermostat\cite{Nose1984, Hoover1985}, while the equations of motion were integrated with a $0.25$ fs timestep. Each cell was relaxed to zero stress along planar directions at the respective temperature via coupling to a thermostat and a barostat. 
Uniaxial strain was applied by increasing the length of the simulation cell along the loading direction at a constant engineering strain rate of $1 \times 10^8$ s$^{-1}$. Even though the atomic positions were not explicitly rescaled in the simulation, the small strain rate prevented the formation of voids, and the atoms moved following their natural dynamics.
The width of the simulation cell, i.e. the length perpendicular to the loading direction, was allowed to relax in order to ensure zero average stress along that direction.
Virial stresses were calculated at each strain level to obtain engineering stress-strain response.


DFT calculations were performed as implemented in the Vienna ab initio simulation package (VASP)\cite{Kresse1996} using the Perdew-Burke-Enzerhof (PBE) generalized gradient approximation exchange-correlation functional \cite{Perdew1996}. The projector augmented wave method was employed with an energy cutoff of 400 eV \cite{Kresse1999}, and a k-point mesh size of $7 \times 7 \times 1$ in the Monkhorst-Pack mesh was employed for Brillouin zone sampling \cite{Monkhorst1976}. A supercell consisting of 72 atoms was fully relaxed with geometry optimization and periodic boundary conditions by the conjugate gradient method. For smearing, the tetrahedron method was applied with a width of $0.1$ eV \cite{Blochl1994a}. After obtaining the optimized structure, we increased the periodic simulation box size in multiple steps along the loading direction with a small engineering strain of 0.0025. In this case to avoid an instant void formation, the atomic positions were rescaled with respect to the box size such that a gap is not formed in the simulation box. A full energy minimization was performed at each step. We studied the mechanical properties under biaxial and uniaxial loading conditions. In the case of biaxial loading, the box size along the perpendicular direction remained constant which resulted in the biaxial stress condition. On the other hand, we allowed the change of simulation box size in perpendicular direction to reach negligible stress to ensure uniaxial stress conditions. The stresses on the minimized structures were measured and plotted against the strain in order to obtain the elastic properties. For the density of state calculation the smearing width was set to $0.01$ eV and a denser k-point mesh size of $11 \times 11 \times 1$ was used. For the ab initio molecular dynamics simulations (AIMD), a Langevin thermostat was used for maintaining the temperature and the integration time step was set at $1.0$ fs.

\begin{figure}[htbp]
\begin{center}
\includegraphics[width=\linewidth]{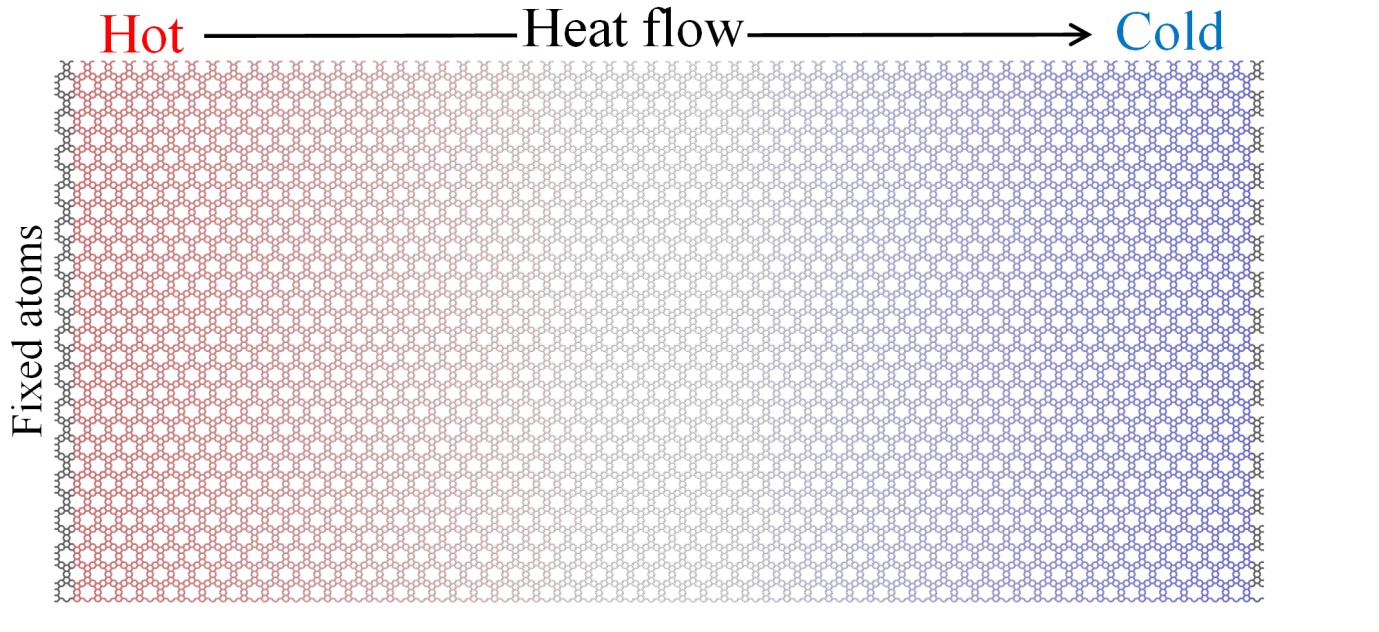}
\caption{System setup for NEMD thermal conductivity calculations. A temperature gradient is imposed on the system by independent thermostats in contact with hot and cold ends of the sample, giving rise to a heat flux.}
\label{fig:nemd}
\end{center}
\end{figure}

The thermal conductivity of NHG was calculated via non-equilibrium molecular dynamics (NEMD) simulations with periodic boundary conditions, in order to investigate size (length) and temperature dependence of heat conduction in this material.
Within this methodology, a temperature gradient is imposed in the system and the resulting heat flux can be  measured, such that the thermal conductivity is directly calculated from Fourier law~\cite{Schelling2002}. 
During NEMD simulations, the entire system was initially equilibrated at the desired temperature via a Nos\'e-Hoover thermostat for $100$ ps, while the equations of motion were integrated with a $0.25$ fs timestep, excluding  fixed atoms at the two ends, as shown in Fig. \ref{fig:nemd}.
Then the thermostat was turned off, and the simulation cell was divided in $21$ slabs along the heat transport direction. The first and last slabs were assigned to be the hot and cold reservoirs, respectively, and were coupled  to independent thermostats, set at different temperatures, while the intermediate slabs were not directly coupled to any thermostats.
The temperature of each slab can be computed from the equipartition theorem as
\begin{equation}
T_i = \frac{2}{3 N_i k_B} \sum_{j} \frac{p_j^2}{2 m_j},
\end{equation}
where $T_i$ is the temperature of $i$-th slab, $N_i$ is the number of atoms in $i$-th slab, $k_B$ is Boltzmann's constant, $m_j$ and $p_j$ are atomic mass and momentum of atom $j$, respectively.
At each simulation step a given amount of energy was added to the atoms in the hot reservoir by the thermostat and an equivalent amount of energy was removed from the atoms in the cold reservoir by the other thermostat, such that on average the total energy exchanged between the system and the thermostats is zero, and the average system temperature is stable. 
The heat flux is given by the difference between the energy added and the energy removed from the system. 
After an initial transient time of approximately $200$ ps the heat flux reaches a stationary state and a stable temperature gradient is established along the simulation cell.
At this stage, the simulation was carried out for at least another $2.0$ ns to calculate the thermal conductivity from the average heat flux and the average temperature gradient. 

\begin{figure}[htbp]
\begin{center}
\includegraphics[width=\linewidth]{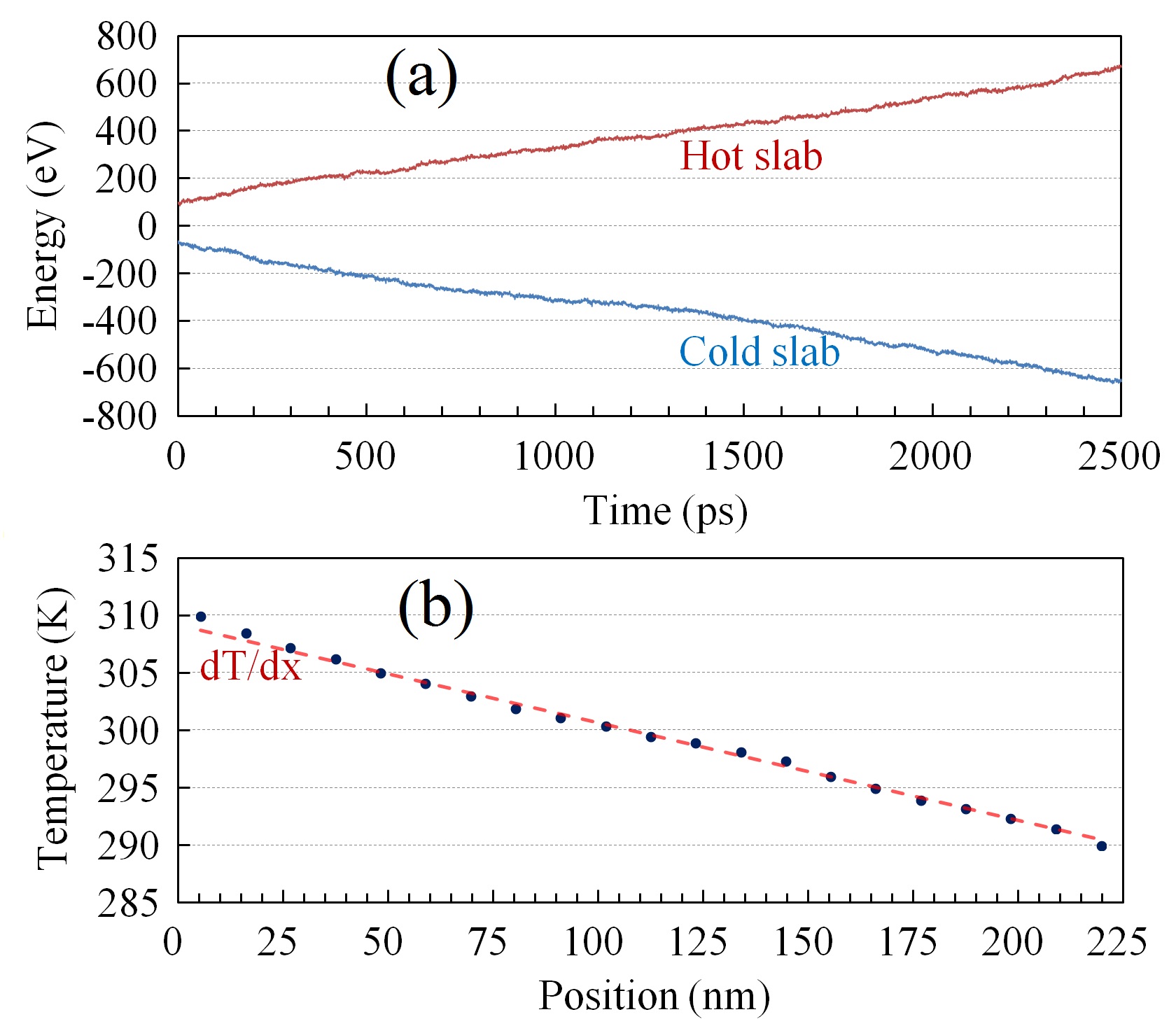}
\caption{(a) Energy added to the hot slab and removed from the cold slab as a function of time. (b) Temperature profile in the system after transient period. Notice the linear profile in the region between hot and cold reservoirs.}
\label{fig:heatflux}
\end{center}
\end{figure}

Fig. \ref{fig:heatflux}(a) shows the energy added to the hot slab and removed from the cold slab as a function of time. The net amount of energy exchanged with the reservoirs averages to zero over time, and thus the total energy is conserved and the system is in a stationary state. In addition, the linear dependence with time, shows that energy is exchanged  at constant rate. The heat flux along the $x$-direction is given by $J_x = \frac{dE/dt}{A}$, where $A$ is the cross-sectional area of the sample given by its width multiplied by thickness. We assumed a thickness of $0.33$ nm for nitrogenated holey graphene which is based on the thickness of graphitic carbon nitride ~\cite{Algara-Siller2014, Mortazavi2015b}.
Fig. \ref{fig:heatflux}(b) shows the time-averaged temperature at each slab. In this case, by neglecting the initial jumps adjacent to the hot and cold slabs, a linear temperature gradient ($dT/dx$) is established along the sample. 
Finally, the thermal conductivity of a sample with size $L$, $\kappa_L$ was computed from the ratio between the average heat flux in a given direction $\langle J_x \rangle$ and the average temperature gradient along the same direction 
\begin{equation}
\kappa_L =  \frac{ \langle J_{x} \rangle}{\langle dT/dx \rangle}.
\label{eq:fourier}
\end{equation}

\section{Results and discussion}

\subsection{Mechanical properties}

\begin{figure}[htbp]
\begin{center}
\includegraphics[width=\linewidth]{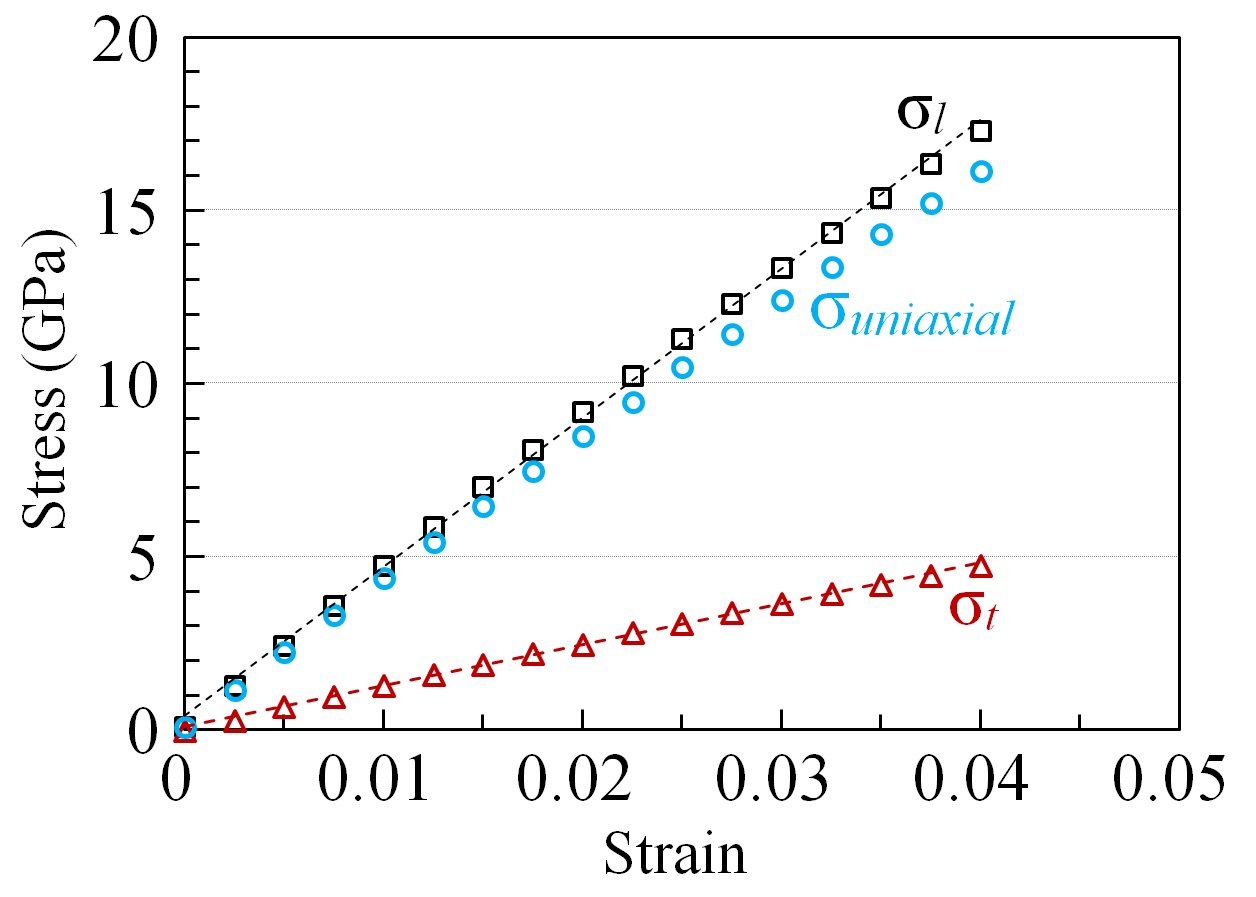}
\caption{First principles density functional theory results of stress-strain relation under uniaxial and biaxial loading conditions. $\sigma_l$ and $\sigma_t$ represent stress values for a  system under biaxial stress along longitudinal and transverse directions, respectively.  $\sigma_{uniaxial}$ is the longitudinal stress for a sample deformed under uniaxial stress. The results correspond to a sample at $0$ K.}
\label{fig:ss}
\end{center}
\end{figure}

In Fig. \ref{fig:ss}, the DFT results for stress-strain response of NHG under uniaxial and biaxial loading is shown. As described above, unidirectional strain was applied along the longitudinal direction while on the transverse direction the box size was fixed. In this case, the stress along the longitudinal ($\sigma_l$) and transverse ($\sigma_t$) directions were calculated at each strain level in order to evaluate elastic properties. As shown in Fig. \ref{fig:ss}, the stress presents a linear dependence with the strain, which confirms that the specimen is stretched within its elastic limit. The Poisson ratio $\nu$, can be obtained using the following ratio:
\begin{equation}
\nu = \frac{\sigma_t}{\sigma_l}.
\end{equation}
Finally, the elastic modulus $E$, can be obtained by:
\begin{equation}
E = \frac{\sigma_l - \nu \sigma_t}{\epsilon},
\end{equation}
where $\epsilon$ is the applied strain. 
Based on our DFT calculations, the elastic modulus under biaxial loading is $398 \pm 5$ GPa, and under uniaxial loading it is  $401.5 \pm 5$ GPa. 
Therefore, the elastic modulus and Poisson ratio of NHG at $0$ K are found to be $400 \pm 5$ GPa and $0.27$, respectively.

\begin{figure}[htbp]
\begin{center}
\includegraphics[width=\linewidth]{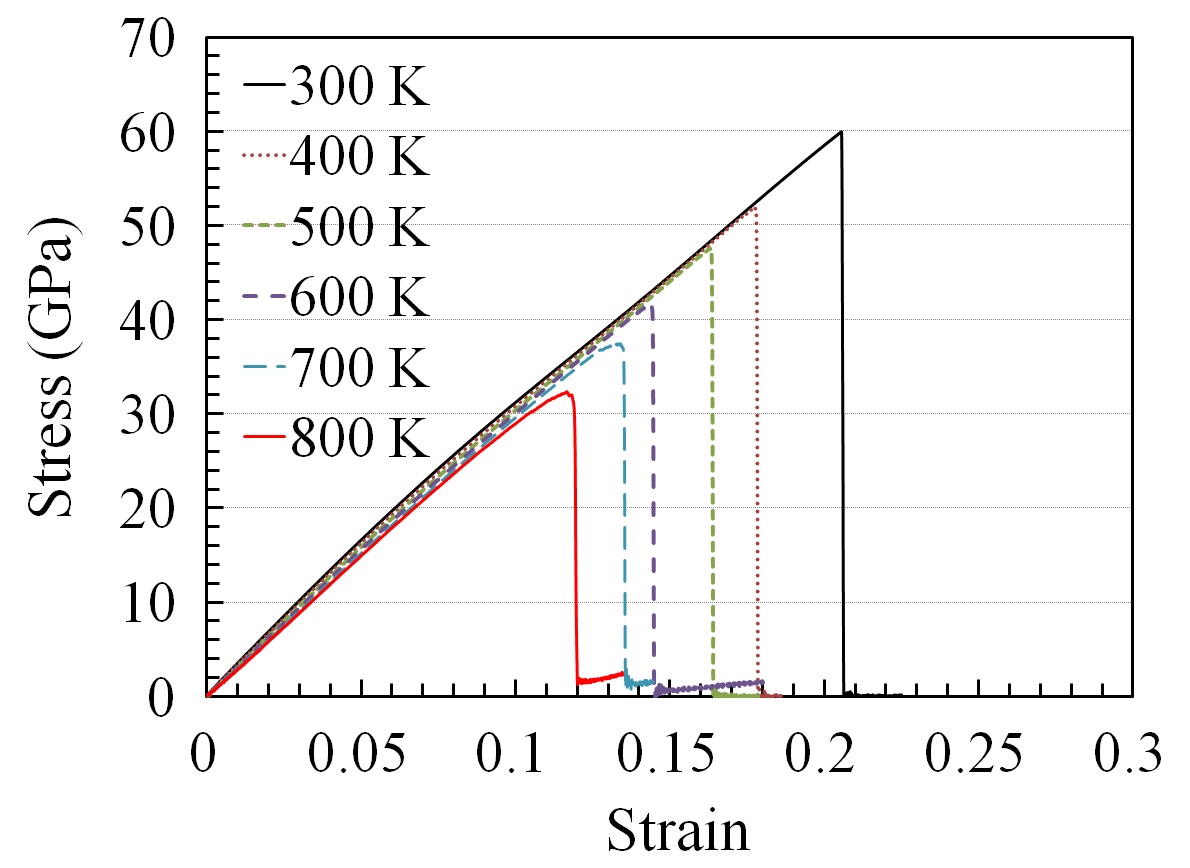}
\caption{Molecular dynamics results for the temperature effects on the stress-strain response of NHG.}
\label{fig:sst}
\end{center}
\end{figure}

In the next step, we performed MD simulations of uniaxial tension at different temperatures. To compare the elastic modulus predicted by MD with that obtained by DFT, we performed the MD modeling of tension at $2$ K. In this case we found an elastic modulus of around $360 \pm 5$ GPa for NHG, which corresponds to a $10\%$ difference from our prediction based on DFT. 
This agreement further validates our choice of parameters for the Tersoff potential and confirms the validity of our MD modeling even though the parameterizations were not trained for NHG. 
In addition, it should be noted that the Tersoff potentials proposed by Lindsay and Broido  were developed to simulate heat conduction along graphene and hexagonal boron-nitride layers and not to exactly reproduce the elastic constants \cite{Lindsay2010, Lindsay2011a}. Furthermore, it is worthy to note that Tersoff potential developed by Lindsay and Broido slightly underestimate (around $3\%$) the elastic modulus of graphene \cite{Mortazavi2014a}. 
Fig. \ref{fig:sst}, present stress-strain relations for NHG at various temperatures from $300$ K to $800$ K. At room temperature, our MD simulations predict an elastic modulus of $335 \pm 5$ GPa and tensile strength of $60$ GPa at corresponding failure strain of $0.206$.
These values are comparable to single-layer graphitic carbon nitride structures \cite{Mortazavi2015b}.
Our results show overall linear decreasing trends in elastic modulus and tensile strength as the temperature increases. This way, at $800$ K the elastic modulus and tensile strength are found to be around $300$ GPa and $32$ GPa, respectively. 

\begin{figure*}[htbp]
\begin{center}
\includegraphics[width=\linewidth]{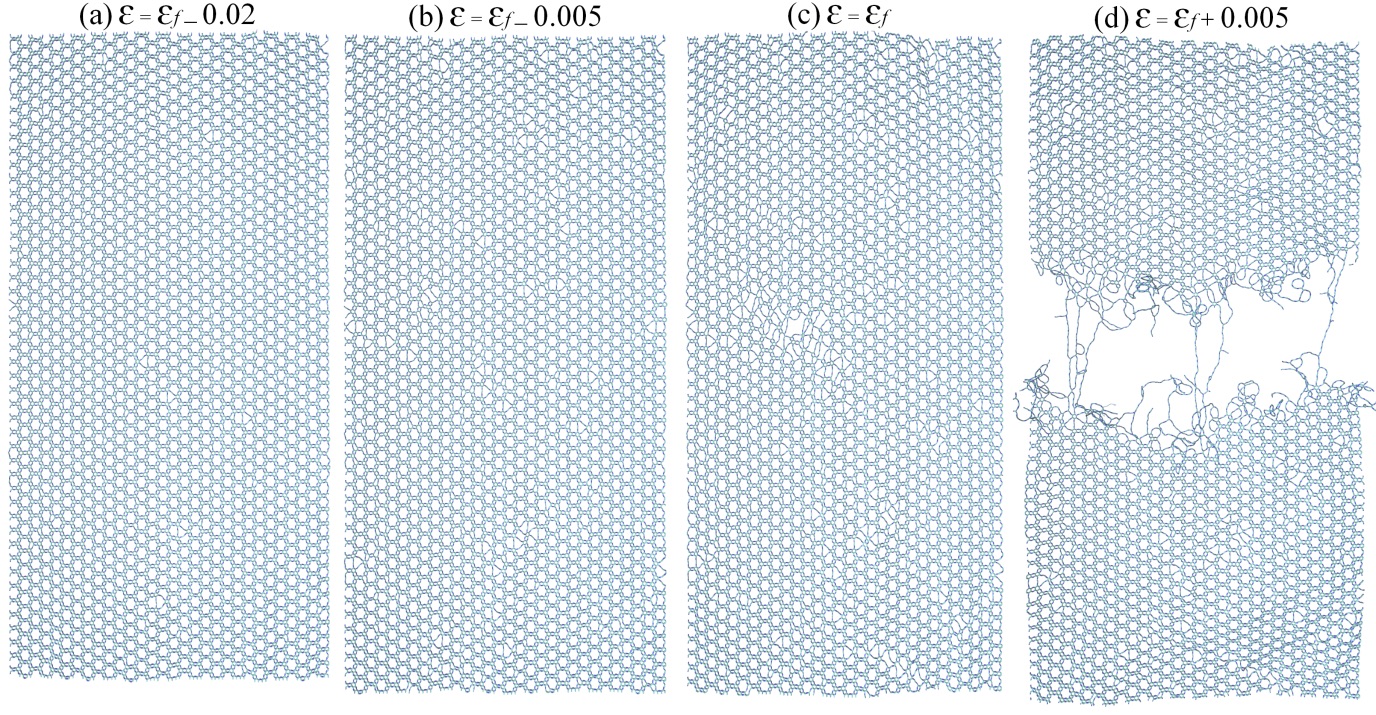}
\caption{Failure process of NHG at $800$ K at different engineering strain levels with respect to the strain at failure, $\varepsilon_f$.}
\label{fig:breaking}
\end{center}
\end{figure*}

The deformation process of a NHG membrane at $800$ K is depicted in Fig. \ref{fig:breaking}. It is found that at strain levels close to the tensile strength, several defects form along the structure. These defects are mainly initiated by breaking of C-C bonds perpendicular to the loading direction. This initial bond breakage instantly cause breaking of adjacent C-C bonds resulting in the formation of large triangular voids consisting of monoatomic chains of carbon and nitrogen atoms. By increasing the strain, more triangular defects form mostly close to the existing defects throughout the sheet ( Fig. \ref{fig:breaking}(b) ). The tensile strength is then found to be a point in which the coalescence of existing defects occur ( Fig. \ref{fig:breaking}(c) ) leading to the sample rupture (Fig. \ref{fig:breaking}(d) ). At high temperatures, the formation of initial defects are slightly earlier than the tensile strength point. In this case, as the initial defects form, the load bearing ability of the structure is reduced. This crack initiation could be realized from the stress-strain response which is the point that the stress start to present non-linear response. Nevertheless, we found that by decreasing the loading temperature the crack initiation and tensile strength occur at closer strain levels. In this regard, at room temperature the crack initiation happens at the tensile strength point and the stress-strain relation follow almost linear pattern up to the tensile strength.  

\subsection{Electronic properties}

\begin{figure}[htbp]
\begin{center}
\includegraphics[width=\linewidth]{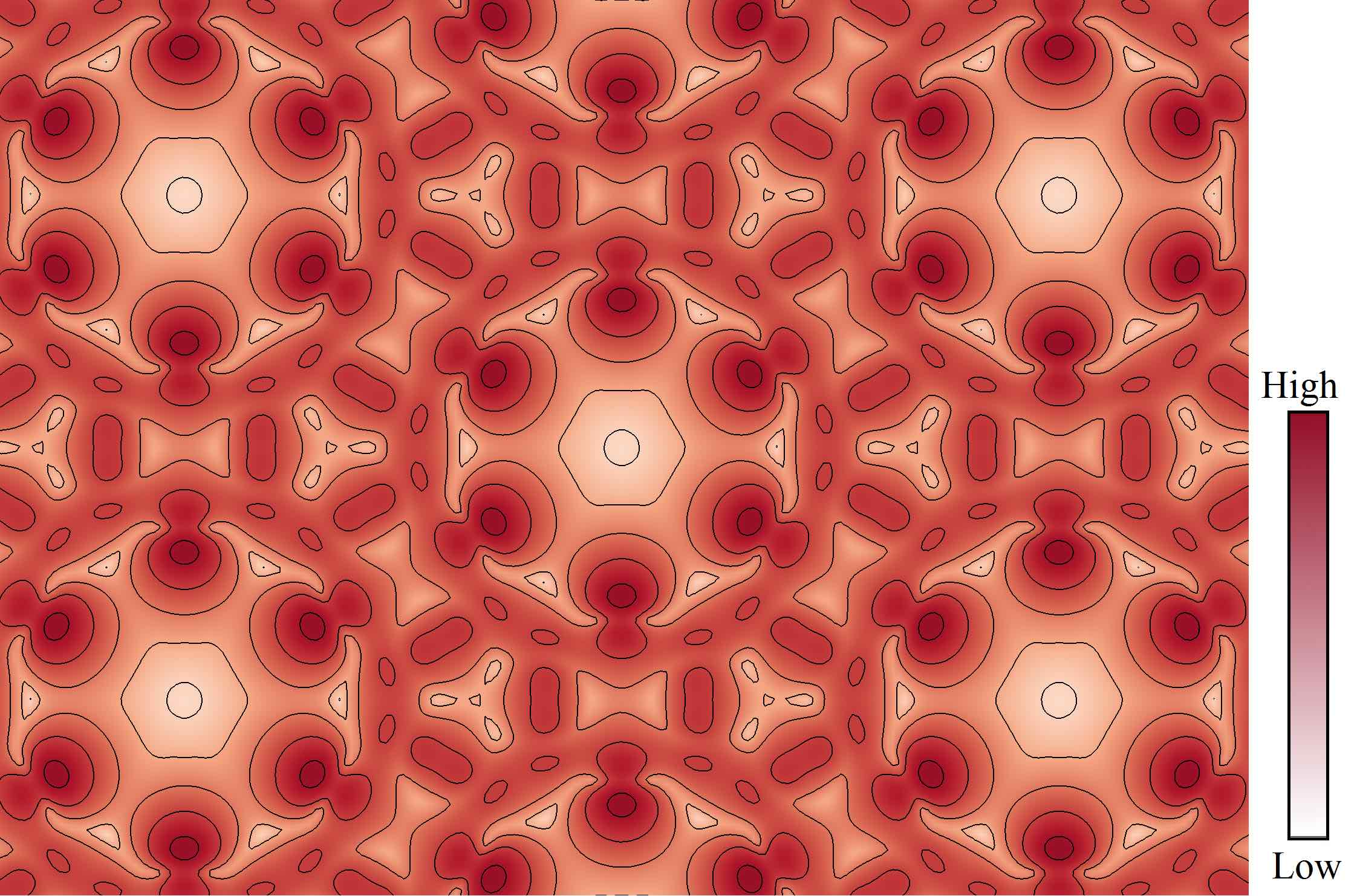}
\caption{Partial electron charge density of free-standing NHG.}
\label{fig:cdens}
\end{center}
\end{figure}

Fig. \ref{fig:cdens}, shows the partial electron charge density distribution of NHG.  Electrons with energies ranging from $-1$ eV up to the Fermi energy were included in the estimation of charge density. Our results illustrate higher concentration of electron charge on the more electronegative nitrogen atoms, especially the lone pairs of electrons are clearly visible. It is found that not all C-C bonds are equivalent as seen from the different distributions of electron charges on them. For half of the C-C bonds, the electron charge is concentrated at the center of the bonds. While for the other half, the electron charge is concentrated on both carbon atoms. These two types are regularly distributed depending on the geometry of the neighboring atoms.  

\begin{figure}[htbp]
\begin{center}
\includegraphics[width=\linewidth]{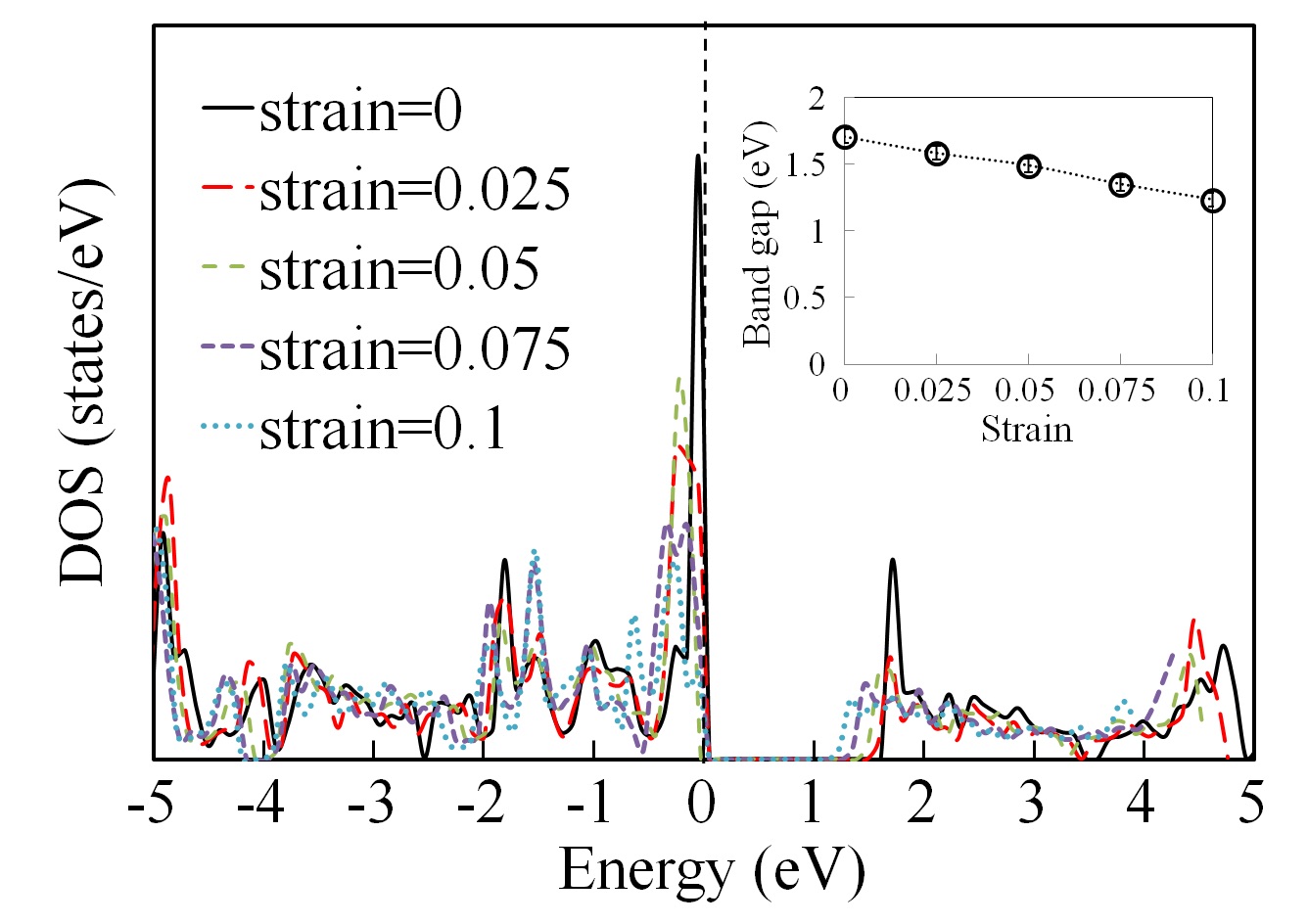}
\caption{Electronic density of states of NHG under uniaxial strain. The inset shows a decrease of the energy gap with strain.}
\label{fig:gap}
\end{center}
\end{figure}

NHG is a semiconductor with a finite band gap of $1.96$ eV as determined optically or $1.7$ eV as determined by DFT \cite{Mahmood2015}. Fig.~\ref{fig:gap} shows the evolution of the electronic density of states and the energy band gap with uniaxial strain as calculated in this work. At completely relaxed state (strain=$0.00$) the band gap was found to be $1.71$ eV, in agreement with the previous DFT study mentioned above. Our first principles calculations show that the band gap decreases monotonically with increasing strain. These results suggest that the band gap of NHG can be manipulated via mechanical strain.

\subsection{Point defects and thermal stability}

\begin{figure*}[htbp]
\begin{center}
\includegraphics[width=\linewidth]{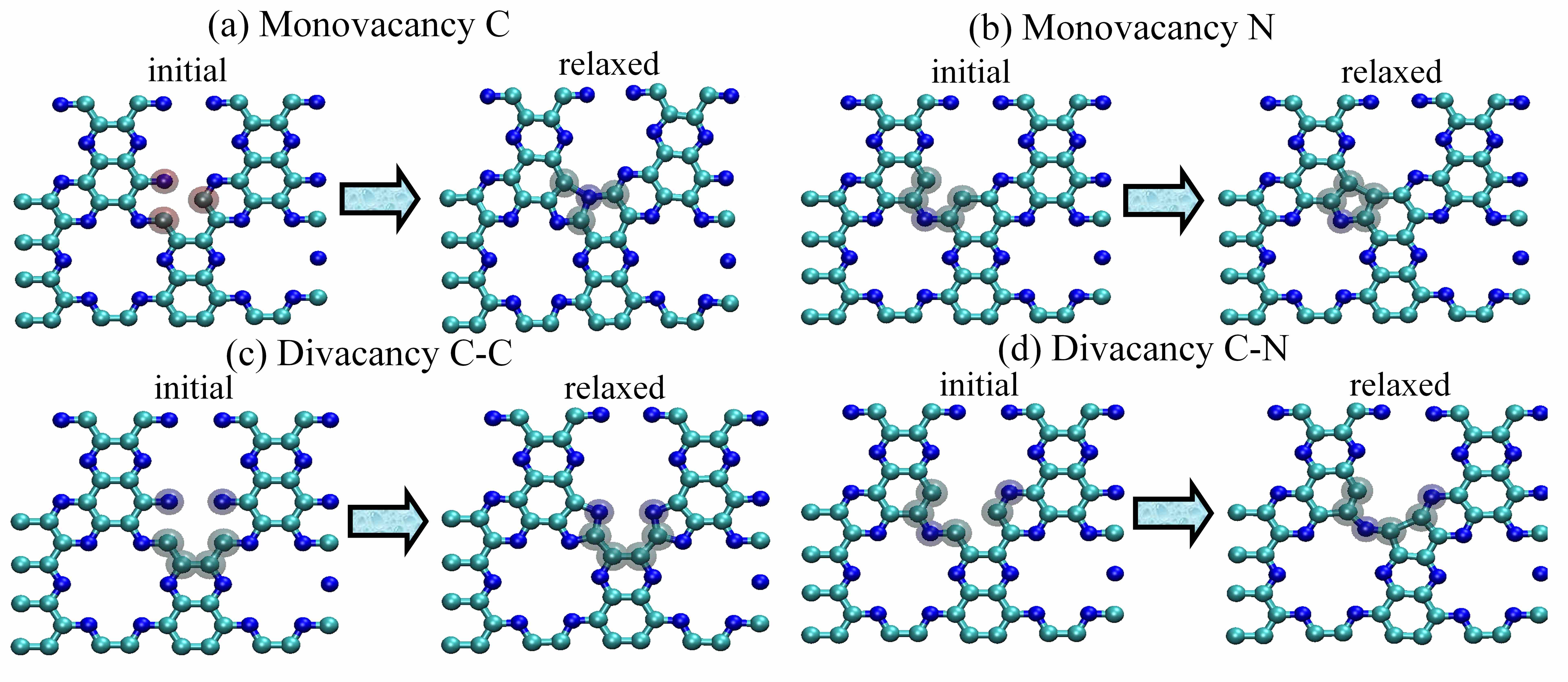}
\caption{Local defects in NHG. Both the initial and relaxed structures are shown. The atoms around the defects are marked.}
\label{fig:defects}
\end{center}
\end{figure*}

In addition to insights from classical MD simulations, we employed DFT to further investigate the stability of NHG against point defects. A supercell of 72 atoms was used for this calculation. We investigated four point defects which included two cases of monovacancy by removing a C atom (case a) or a N atom (case b), and two cases of divacancies where we removed a pair of C atoms (case c) or one C and one N atom (case d). After introducing the defect, the geometry of each structure was relaxed via energy minimization. Fig.~\ref{fig:defects} presents the energy-minimized configurations, and shows that the introduction of defects does not change the global structure of NHG. During minimization, new bonds were formed with neighboring atoms in order to compensate for missing atoms. The overall distortions caused by the defects are local by nature, which confirms the robustness and stability of NHG's structure against defects.

Next, the thermal stability of the defective structures was also tested via AIMD simulations at $300$ K. Each structure was simulated for a total time of $5.0$ ps, and the defective structures remained intact at the end of each simulation.  
Comparing the potential energies of each structure during AIMD simulations we concluded that divacant structures  are less stable than monovacant structures. We also noticed that the monovacant C structure is less stable than the monovacant N structure, and that the divacant CC structure is less stable than the divacant CN structure.

To further elaborate on the role of defects in the thermal stability of NHG we conducted AIMD simulations at very high temperatures, namely $T=3000$ K and $T=4000$ K for pristine and defective NHG. These simulations were also  conducted for $5.0$ ps. At the end of each simulation pristine NHG and all four defective structures were intact at $T=3000$ K. This suggests high thermal stability of NHG in spite of local defects. At $T=4000$ K the pristine NHG structure was partially disintegrated at the end of the simulation, and all of the four defective structures were completely disintegrated at this temperature. This shows that point defects can play a role in initiating thermal disintegration of NHG at very high temperatures.  
Thus, we conclude that the point defects considered in this study only affect NHG locally. Nonetheless, at a very high temperature these defects can play a role in disturbing its structural integrity and thermal stability.   

\subsection{Thermal conductivity}

\begin{figure}[htbp]
\begin{center}
\includegraphics[width=\linewidth]{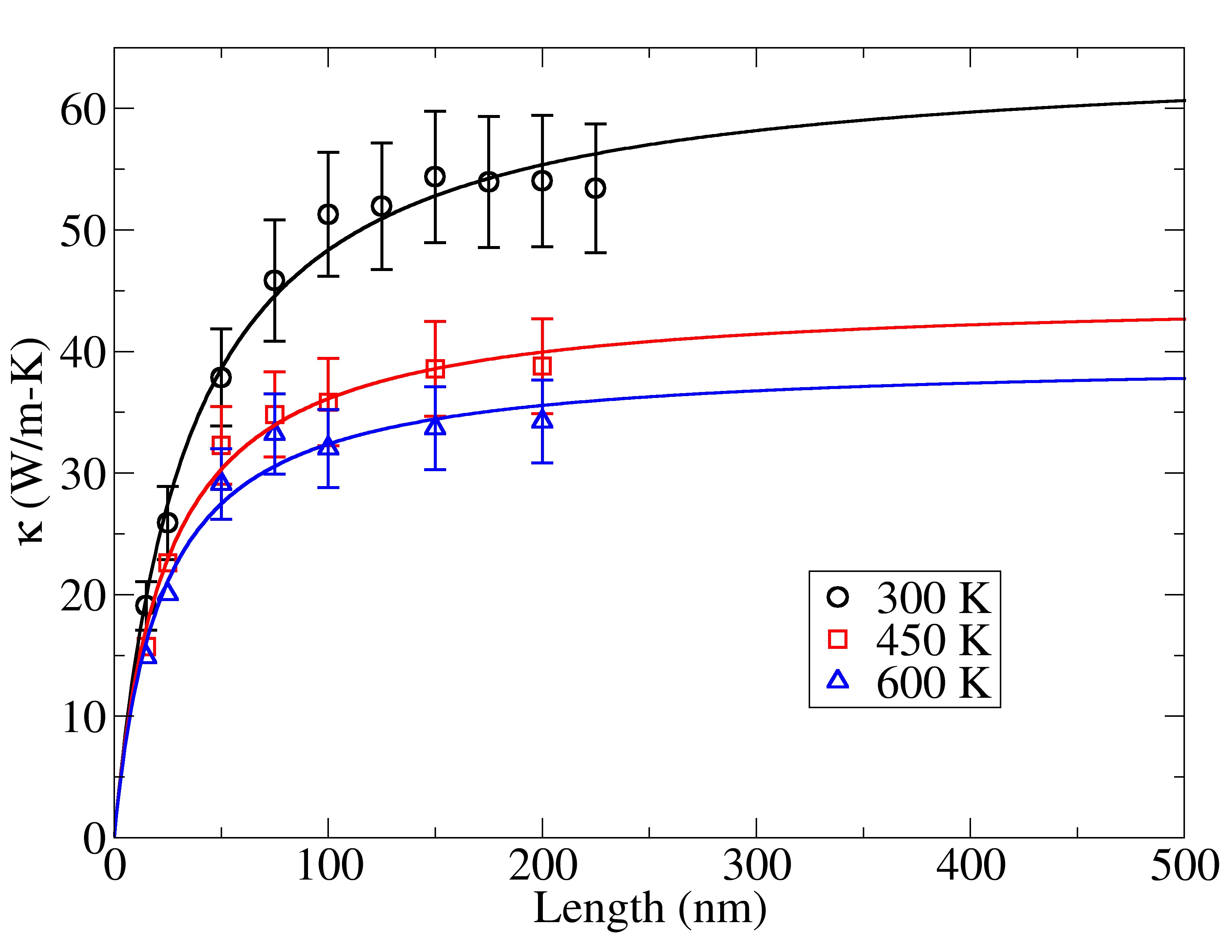}
\caption{Length dependence of NHG's thermal conductivity. Solid lines are best fit to Eq. \ref{eq:kL}, with parameters given in Table~\ref{table:I}.}
\label{fig:cond-vs-L}
\end{center}
\end{figure}

Thermal conductivity calculations with the NEMD method present strong size dependence effects~\cite{Xu2014}. In order to investigate this size dependence in NHG we performed simulations with periodic cells of increasing length and a constant width of $23$ nm, as shown by the data points in Fig.~\ref{fig:cond-vs-L} for three temperatures.
For very small simulation cells $\kappa_L$ increases linearly with sample length, which corresponds to the ballistic transport limit. In NHG, for large enough cells the conductivity converges to a finite value, corresponding to the diffusive regime. At intermediate lengths, the system is in a cross-over region from ballistic to diffusive heat transport.
In general, the dependence of $\kappa$ on $L$ can be described by \cite{Schelling2002}
\begin{equation}
\frac{1}{\kappa_L} = \frac{1}{\kappa_{\infty}} \left( 1+ \frac{\Lambda}{L} \right),
\label{eq:kL}
\end{equation}
where $\kappa_{\infty}$ is the size-independent thermal conductivity of the material (i.e. the conductivity in the diffusive regime), and $\Lambda$ is an effective phonon mean free path (MFP) for the material.
From the equation we see that $L=\Lambda$ yields $\kappa_L=\kappa_{\infty}/2$, such that this effective MFP corresponds to the system length at which the system reaches $50 \%$ of its diffusive thermal conductivity \cite{Minnich2011, Zhang2015, Neogi2015}.
Therefore, we can simultaneously estimate the intrinsic diffusive thermal conductivity of a given material as well as its effective phonon MFP via Eq.\ref{eq:kL}.
The solid lines in Fig.~\ref{fig:cond-vs-L} represent the best fit to Eq.~\ref{eq:kL} for each temperature. The values of $\kappa_{\infty}$ and $\Lambda$ for each temperature are summarized in Table~\ref{table:I}.

We predict a thermal conductivity of $64.8$ W/m-K for NHG at room temperature, which is two orders of magnitude smaller than the thermal conductivity of graphene  at the same temperature \cite{Ghosh2010, Xu2014, Fugallo2014}.
The effective phonon MFP of isotopically pure graphene at $300$ K can be estimated to be $\approx 5$ $\mu$m \cite{Fugallo2014}, also two orders of magnitude longer than the calculated $\Lambda=34.0$ nm for NHG.
We attribute such difference mostly to the lower density of NHG relative to graphene, as well as the presence of N atoms in the structure which soften the phonon modes of pristine graphene.
Nonetheless, it is interesting to notice that the problem of collective excitations with very long MFP does not materialize in NHG \cite{Fugallo2014}, even near room temperature.
Indeed, from Fig.~\ref{fig:cond-vs-L} we observe that $\kappa$ seems to converge at a sample length of $\approx 150$ nm even at room temperature.
It is also interesting to compare the effective phonon MFP with the characteristic length scales of this crystal structure, namely the diameter of the holes and the distance between them. At room temperature, the average hole diameter equals $0.62$ nm while the average distance between them is $0.88$ nm. 
Therefore, at room temperature, both length scales are much shorter than the effective phonon MFP. 

\begin{table}[htp]
\caption{Intrinsic thermal conductivity and effective phonon mean free path for NHG.}
\begin{center}
\begin{tabular}{c | c | c}
\hline
\hline
T (K) & $\kappa_{\infty}$ (W/m-K) & $\Lambda$ (nm) \\
\hline
300 & 64.8 & 34.0 \\
450 & 44.7 & 23.8 \\
600 & 39.4 & 21.8 \\
\hline
\hline
\end{tabular}
\end{center}
\label{table:I}
\end{table}%

\begin{figure}[htbp]
\begin{center}
\includegraphics[width=\linewidth]{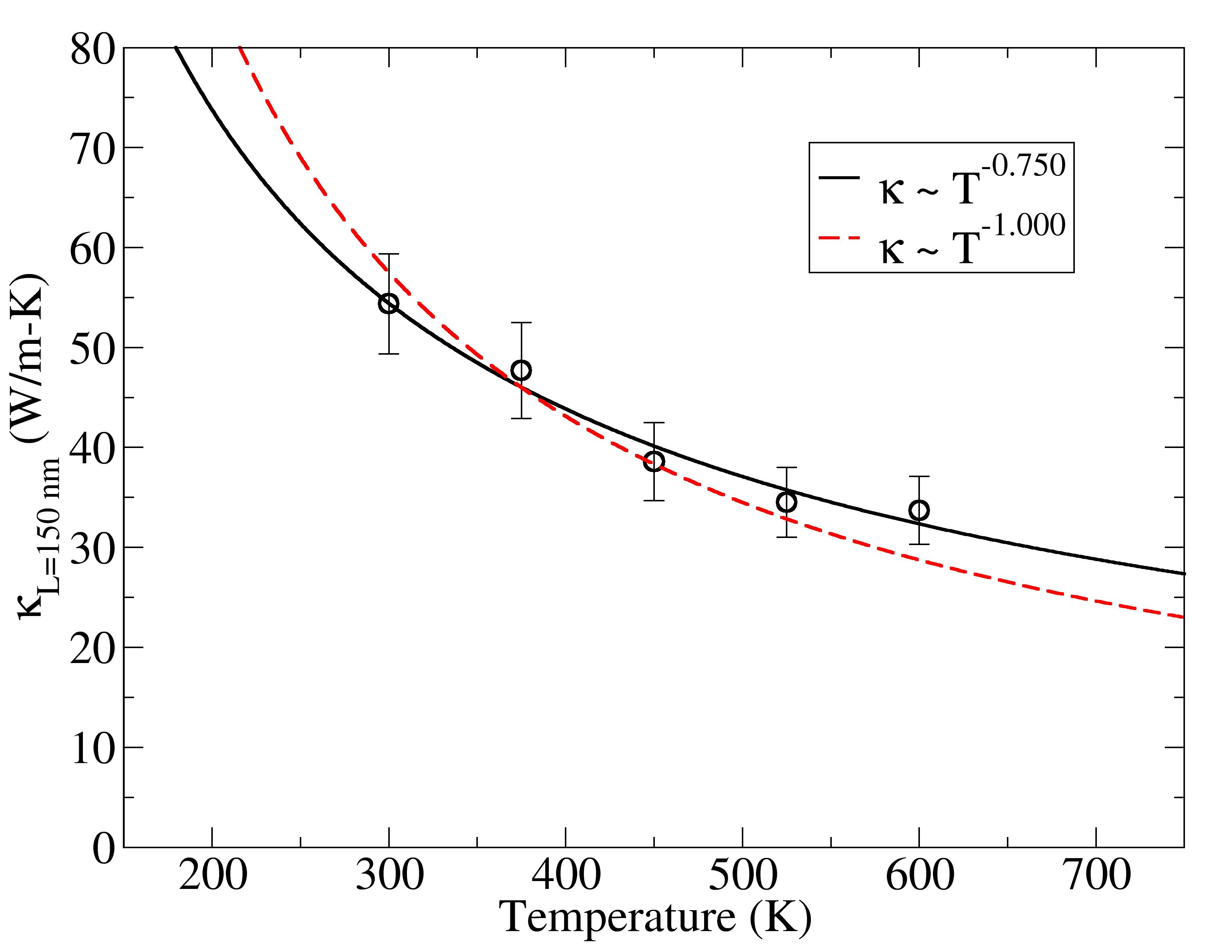}
\caption{Temperature dependence of NHG's thermal conductivity at a fixed length $L=150$ nm. Deviation from  inverse temperature trend is attributed to dimensionality of NHG, the presence of holes and two different atomic species in the material.}
\label{fig:condT}
\end{center}
\end{figure}

From Table \ref{table:I} we observe a decrease of both $\kappa$ and $\Lambda$ as temperature increases, as qualitatively expected. 
In Fig. \ref{fig:condT} we present the thermal conductivity for samples with a fixed length $L=150$ nm, as a function of simulation temperature, for temperatures above $300$ K.
In this temperature range, and in the absence of structural defects and impurities, the thermal conductivity of a bulk crystalline material is expected to be inversely proportional to temperature~\cite{Ziman1960, Pereira2013}. In other words, when the only source of thermal resistance is phonon-phonon scattering then $\kappa \sim T^{-1}$. However, in the presence of defects and impurities, $\kappa$ becomes less dependent on $T$, and the exponent assumes a smaller (absolute) value.
When phonon-defect scattering dominates over phonon-phonon scattering, the conductivity is expected to be temperature-independent, such that one can write $\kappa \sim T^{0}$~\cite{Ziman1960, Donadio2010}.
Fig. \ref{fig:condT} shows that our data agrees with an exponent smaller than unit, a behavior we attribute to the dimensionality of NHG, as well as to the presence of holes and two different atomic species in the material.
Based on our AIMD simulations, we believe that no structural defects are formed in our MD simulations.
Nonetheless, the scattering rate of phonons by lattice holes increases with temperature, leading to higher thermal resistance and lower thermal conductivity.
We have recently performed a study of mechanical properties and thermal conductivity of amorphized graphene samples, in which we found that certain types of defects produce a larger decrease in thermal conductivity ~\cite{Mortazavi2016}. On the same study we observed that, for samples with high defect concentration, thermal conductivity is almost temperature-independent.
A full characterization of structural defects and their individual effect on the thermal conductivity of NHG shall be addressed in future work.

\section{Conclusions}

In conclusion, we performed first principles calculations and extensive atomistic MD simulations to investigate mechanical properties and thermal conductivity of NHG, a novel two-dimensional material, at various temperatures.
MD simulations were based on the Tersoff potential with parameters obtained from graphene and hBN parameterizations.
The phonon dispersion of NHG presents no instabilities with the chosen empirical potential.
Mechanical properties were evaluated via uniaxial tensile strain  to obtain elastic constants.
First principles DFT calculations predict an elastic modulus of $400 \pm 5$ GPa at $0$ K, which is $10\%$ larger than the value obtained by atomistic MD simulations at very low temperatures.
Results show an overall decreasing trend in elastic modulus and tensile strength as temperature increases. 
At room temperature, our MD simulations predict an elastic modulus of $335 \pm 5$ GPa and a tensile strength of $60$ GPa at a corresponding failure strain of $0.206$.
During mechanical failure under strain we observed defect initiation by breakage of C-C bonds mainly perpendicular to the loading direction, formation of large triangular voids consisting of mono-atomic chains and finally void coalescence leading to the final rupture of the sample.
We also investigate the thermal conductivity of NHG via NEMD simulations, in particular its length dependence.
Fitting simulation results to an empirical expression for $\kappa$ we predict an intrinsic thermal conductivity of $64.8$ W/m-K and a corresponding effective phonon mean free path of $34.0$ nm at $300$ K.
Both limiting thermal conductivity and effective phonon mean free path decay with temperature for temperatures above $300$ K.
Finally, due to the current lack of experimental studies on the physical properties of NHG, our modeling-based predictions should serve as a guide to future experiments concerning the physical properties of this novel 2D material.

\section*{Acknowledgments}
B.M. and T.R. greatly acknowledge the financial support by European Research Council for COMBAT project.
L.F.C.P. would like to thank Hasan Sahin and Mehmet Yagmurcukardes for providing DFT-optimized unit cell structure for comparison of phonon dispersions.  
L.F.C.P. acknowledges financial support from Brazilian government agency CAPES for project ``Physical properties of nanostructured materials" via its Science Without Borders program.

\section*{References}

\bibliography{/Users/pereira/Dropbox/Documents/library}

\end{document}